\documentclass[runningheads]{llncs}
\usepackage[T1]{fontenc}
\usepackage{graphicx,verbatim}
\usepackage{hyperref}
\usepackage{amsmath}
\usepackage{cleveref}
\usepackage[inline]{enumitem}
\usepackage{tabularx}
\usepackage{booktabs}
\usepackage{multirow}
\usepackage{makecell}
\usepackage{xcolor,colortbl}
\usepackage{float}
\usepackage{placeins}
\usepackage{orcidlink}

\usepackage{color}

\begin{document}

\title{The Missing Piece: A Case for Pre-Training in 3D Medical Object Detection}
\author{Katharina Eckstein\inst{\star,1,2,3}\orcidlink{0009-0009-4038-3679}\and
Constantin Ulrich\inst{\star,1,2}\orcidlink{0000-0003-3002-8170}\and \\
Michael Baumgartner\inst{\star\star,1,4,5}\orcidlink{0000-0003-4455-9917}\and
Jessica Kächele\inst{1,2,3}\orcidlink{0000-0003-3851-2848}\and
Dimitrios Bounias\inst{1,2}\orcidlink{0000-0002-3361-1698}\and
Tassilo Wald\inst{1,4,5}\orcidlink{0009-0007-5222-2683}\and
Ralf Floca\inst{1,6}\orcidlink{0000-0003-3218-3377}\and
Klaus H. Maier-Hein\inst{1,2,3,4,5,7,8}\orcidlink{0000-0002-6626-2463}
}
%index{Eckstein, Katharina}
%index{Ulrich, Constantin}
%index{Baumgartner, Michael}
%index{Kächele, Jessica}
%index{Bounias, Dimitrios}
%index{Wald, Tassilo}
%index{Floca, Ralf}
%index{Maier-Hein, Klaus H.}
\authorrunning{K. Eckstein \& C. Ulrich et al.}
\institute{
Medical Faculty Heidelberg, Heidelberg University, Heidelberg, Germany\and
German Cancer Consortium (DKTK), German Cancer Research Center (DKFZ), Core Center Heidelberg, Germany\and
Helmholtz Imaging, DKFZ, Heidelberg, Germany\and
Faculty of Mathematics and Computer Science, Heidelberg University, Germany\and
Heidelberg Institute of Radiation Oncology (HIRO), National Center for Radiation Research in Oncology (NCRO), Heidelberg, Germany\and
Pattern Analysis and Learning Group, Department of Radiation Oncology, Heidelberg University Hospital, Heidelberg, Germany\and
National Center for Tumor Diseases (NCT) Heidelberg, Heidelberg, Germany
\email{\{katharina.eckstein,constantin.ulrich\}@dkfz-heidelberg.de}}
\maketitle
\footnotetext[1]{Equal contribution; co-first author order may be adjusted for individual use.}
\footnotetext[2]{Work done while at DKFZ, now at Siemens Healthineers.}
\begin{abstract}
Large-scale pre-training holds the promise to advance 3D medical object detection, a crucial component of accurate computer-aided diagnosis. Yet, it remains underexplored compared to segmentation, where pre-training has already demonstrated significant benefits. Existing pre-training approaches for 3D object detection rely on 2D medical data or natural image pre-training, failing to fully leverage 3D volumetric information. In this work, we present the first systematic study of how existing pre-training methods can be integrated into state-of-the-art detection architectures, covering both CNNs and Transformers. Our results show that pre-training consistently improves detection performance across various tasks and datasets. Notably, reconstruction-based self-supervised pre-training outperforms supervised pre-training, while contrastive pre-training provides no clear benefit for 3D medical object detection. Our code is publicly available at: \href{https://github.com/MIC-DKFZ/nnDetection-finetuning}{https://github.com/MIC-DKFZ/nnDetection-finetuning}.
\end{abstract}
\section{Introduction}
Accurate detection of anatomical structures and abnormalities in 3D medical imaging is crucial for reliable diagnosis and clinical decision-making. Unlike segmentation, which provides detailed structural delineation, detection focuses on localizing clinically relevant objects. Critically, detection excels in clinically relevant metrics, especially in high-stakes scenarios where completely missing an object can have far more severe consequences than minor inaccuracies in pixel-wise delineation \cite{metrics_reloaded}. Despite its clinical importance, research on 3D object detection has received significantly less attention than segmentation, as evidenced by medical image analysis challenges predominantly emphasizing segmentation tasks \cite{maier2018rankings}.

Recent advancements in 3D medical image segmentation have spurred interest in large-scale pre-training. For instance, Ulrich et al. introduced Multitalent \cite{Ulrich_2023}, a framework that enables supervised training across multiple segmentation datasets. Moreover, self-supervised pre-training strategies \cite{wald2024revisitingmaepretraining3d,ModelsGenesis,VoCo,tian2023designingbertconvolutionalnetworks,he2022masked} have demonstrated promising results for segmentation applications. Likewise, detection models might particularly benefit from pre-training due to the typically small size of annotated datasets and their tendency to over focus on local image features, rather than leveraging broader contextual information. However, despite the advancements in pre-training for segmentation, the impact of purely 3D large-scale pre-training remains unexplored for 3D object detection.

This gap was also acknowledged in one of the most recent and comprehensive studies on 3D medical object detection by Baumgartner et al., who extensively revised the nnDetection framework \cite{Baumgartner_2021,nndetv2}. While their work made significant contributions to the field, it did not address the potential role of large-scale pre-training. Yet, beyond their work, research on pre-training strategies for medical object detection is virtually nonexistent, with only a handful of studies even touching upon this direction. Existing pre-training strategies for medical object detection have predominantly focused on 2D data, utilizing either natural image pre-training \cite{wu2023zero} or 2D medical data \cite{Liu2023MFLAGMV,cheng2023prior,muller2022joint,benvcevic2022self,UtilizingSD}. This is largely due to the scarcity of publicly available 3D object detection datasets with sufficient cases for effective pre-training. To partially capture 3D context, some methods extend pre-trained 2D models by integrating adjacent slices. This includes using ImageNet-pretrained backbones with 3D context slices added at the downstream stage \cite{wang2019volumetric,yan2020learning} or pseudo-3D approaches that treat image channels (e.g., RGB) as separate slices during pre-training \cite{zhang2020revisiting}. Another strategy relies on video-based pre-training, where adjacent frames are used to simulate the sequential nature of medical slices \cite{amiriparian2023universal}. Notably, no prior study has systematically explored large-scale 3D pre-training for 3D medical object detection. 

To bridge this gap, we present a comprehensive study evaluating the impact of different large-scale pre-training strategies on 3D medical object detection. Specifically, our key contributions include:
\begin{enumerate}
    \item \textbf{The First Comprehensive Study on Pre-Training Paradigms for 3D Object Detection} to analyze the impact of both supervised and self-supervised large-scale pre-training for 3D medical object detection across eight diverse downstream detection datasets.
    \item \textbf{Evaluation Across Detection Architectures:} Model performance varies widely depending on dataset characteristics and annotation types, such as bounding boxes or segmentation masks, as demonstrated by Baumgartner et al. \cite{nndetv2}. To assess the generalization of pre-training strategies, we examine their transferability to two state-of-the-art detection models, Retina U-Net \cite{RetinaUNet,nndetv2} and Deformable DETR \cite{DeformableDETR,nndetv2}, covering both CNN and Transformer.
    \item \textbf{Comparison of Pre-Training Architectures for Pre-training:} ResEncL, a state-of-the-art model for semantic segmentation \cite{isensee2024nnu} that has shown improved downstream performance with self-supervised pre-training \cite{wald2024revisitingmaepretraining3d}, and an adapted version of Retina U-Net, allowing both segmentation pre-training as well as downstream 3D object detection fine-tuning.
\end{enumerate}
\section{Methods}
\begin{figure}[t]
    \centering
    \includegraphics[width=\linewidth]{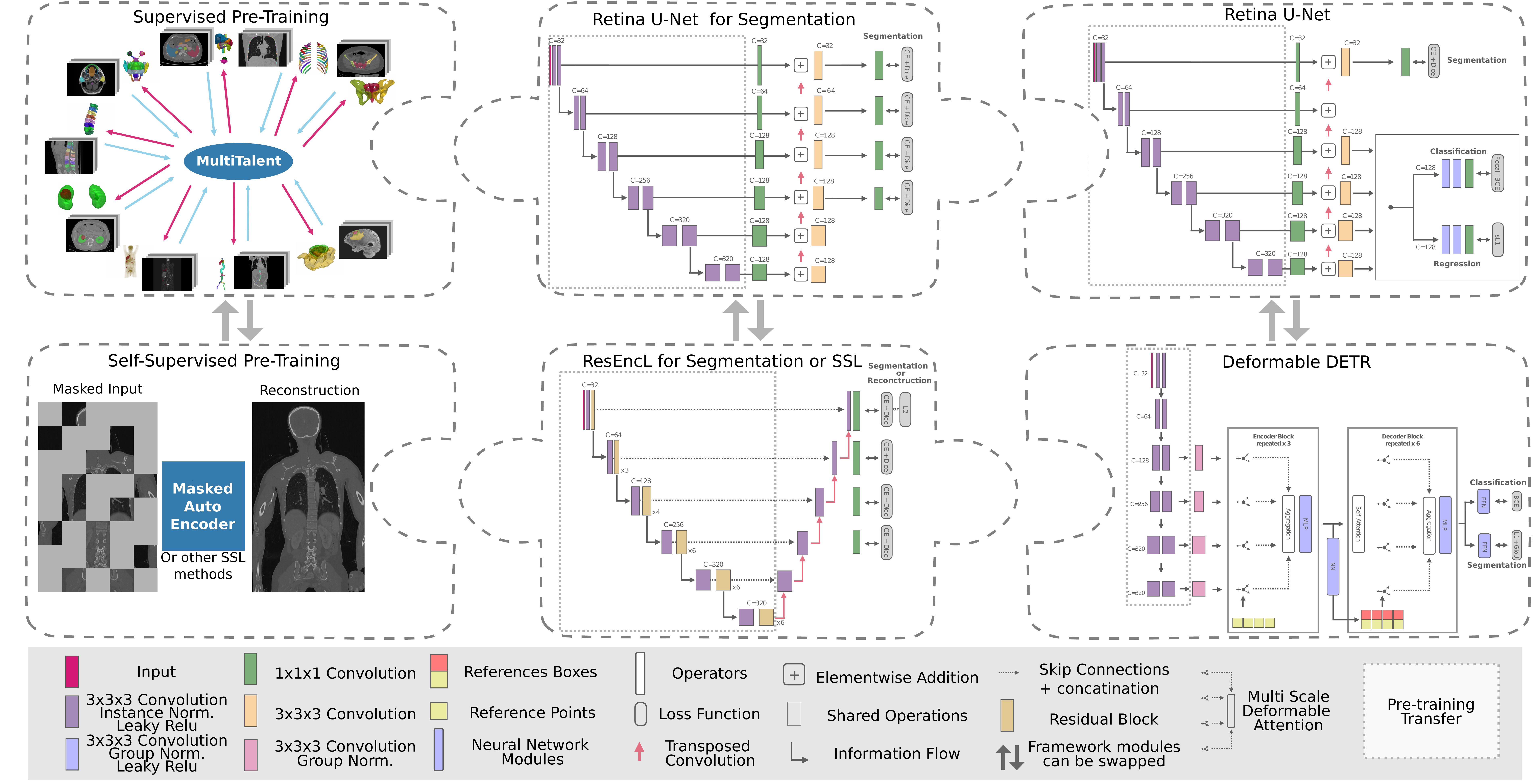}
    \caption{\textbf{A cross-framework bridge between nnDetection and its pre-training counterparts}: Different pre-training paradigms (supervised and self-supervised), pre-training architectures (Retina U-Net and ResEncL), and detection-specific models (Retina U-Net and Deformable DETR) can be combined like puzzle pieces, offering a flexible and integrative approach to optimizing detection performance.}
    \label{fig:framework}
\end{figure}

In this study, we evaluate the impact of large-scale pre-training on 3D medical object detection using two state-of-the-art architectures: Retina U-Net and Deformable DETR. Notably, both architectures are specifically designed for detection and cannot be directly applied to other tasks without modifications. Therefore, for pre-training, we adapt Retina U-Net for supervised segmentation and employ the state-of-the-art ResEncL model for both supervised and self-supervised learning \cite{wald2024revisitingmaepretraining3d,isensee2024nnu}. We then transfer only the pre-trained backbone from these models to the detection networks for downstream fine-tuning, as visualized in \cref{fig:framework}.
Our experimental setup involves five development datasets and three independent testing datasets. The development datasets are employed to systematically investigate various fine-tuning strategies, enabling us to identify optimal approaches for adapting pre-trained models to 3D medical object detection. %The testing datasets are reserved for evaluating the generalizability of these strategies on unseen data.
\subsection{3D Object Detection}
\subsubsection{Retina U-Net}
 \cite{RetinaUNet} is a single-stage, anchor-based object detector enhanced with semantic segmentation supervision. Its architecture extends the Feature Pyramid Network (FPN) of RetinaNet with additional high-resolution levels in the FPN’s top-down pathway to support an auxiliary segmentation task, creating a U-Net-like symmetric structure (U-FPN), as visualized in \cref{fig:framework}. The detection head, applied to the final four or five resolution levels, consists of a classification and a regression branch. The regression branch uses smooth L1 loss, while the classification branch employs binary focal loss. Segmentation is supervised with a combined cross-entropy and batch Dice loss function.

\subsubsection{Deformable DETR}
\cite{DeformableDETR} is a two-stage transformer-based detection architecture. In contrast to traditional DEtection TRansformer (DETR), Deformable DETR replaces global self-attention with a sparse deformable attention mechanism, significantly reducing computational complexity and enhancing efficiency by focusing on a small set of queries per attention operation. Additionally, Deformable DETR introduces iterative bounding box refinement, progressively updating the bounding boxes instead of predicting them from scratch. As visualized in \cref{fig:framework}, Deformable DETR contains an encoder network as a first component to extract a feature representation from the input image. A point-wise convolution is applied to this feature representation to reduce the channel dimensionality. The extracted feature maps are flattened into a sequence of spatial tokens, with positional encodings added to retain spatial information. These tokens are then processed by a Transformer encoder-decoder architecture (3 encoder and 6 decoder blocks). The Deformable DETR detection head comprises one branch for classification (linear layer) and one for bounding box prediction (multi-layer perceptron). Focal loss is employed to account for dataset imbalances. 

\subsection{Pre-training Paradigms}
\subsubsection{Supervised Pre-training}
For large-scale supervised pretraining, we adopted the MultiTalent (MT) approach – a multi-dataset training paradigm introduced by Ulrich et al. \cite{Ulrich_2023}.
To support this, we compiled a large-scale dataset collection of publicly available, pixel-wise annotated 3D medical images, comprising over 20,000 3D volumes sourced from 65 datasets, with more than 300,000 image-mask pairs. The dataset includes CT, MRI, and PET modalities. A more detailed description of the datasets can be found in the appendix (see section \ref{sec:MT_datasets}). Notably, several datasets feature re-annotated publicly available images—for instance, the Abdomen Atlas \cite{abdomenatlas} and Abdomen1K \cite{AbdomenCT-1K} datasets include images from the Medical Decathlon, among others. All datasets and images used for downstream fine-tuning were excluded from pre-training to avoid data leakage.

\subsubsection{Self-Supervised Pre-training}
Self-supervised pre-training was performed using two large-scale medical imaging datasets: CT-RATE \cite{CT-RATE} and the Adolescent Brain Cognitive Development (ABCD) Study \cite{ABCD}, totaling 91,768 training images. CT-RATE includes 25,692 non-contrast 3D CT scans, expanded to 50,188 volumes through multiple reconstructions from 21,304 unique patients.
The ABCD Study, the largest U.S. longitudinal brain development study, contributed 41,580 brain images from 11,875 participants aged 9–10 at baseline, including T1-weighted, T2-weighted, and fMRI scans.
We evaluated four self-supervised pre-training paradigms: \\
\textit{Models Genesis (MG)} aims to reconstruct original image patches from transformed versions using non-linear intensity shifts, in-painting, out-painting, and local shuffling techniques \cite{ModelsGenesis}.\\
\textit{Masked Autoencoder (MAE)} utilizes a masked autoencoding strategy to reconstruct images, applying a 75\% mask ratio to learn contextual features \cite{he2022masked}.\\
\textit{SparkMAE (S3D)} modifies MAEs for CNN architectures to better process sparse inputs. It introduces sparse convolutions and normalization, where masking is reapplied after each convolution and normalization is restricted to non-masked values. A learnable mask token is used to fill masked areas for the encoder, followed by a densification convolution layer applied to all but the highest resolution feature maps \cite{tian2023designingbertconvolutionalnetworks}.\\
\textit{VoCo} leverages anatomical consistency by contrasting random sub-volumes\\
against base crops to predict contextual overlap within 3D medical images \cite{VoCo}. 

\subsubsection{Implementation}
We trained two MultiTalent networks: the state-of-the-art segmentation model ResEncL U-Net \cite{isensee2024nnu}, and Retina U-Net \cite{RetinaUNet}. The ResEncL U-Net employed a patch size of cubic 192 with a batch size of 12, while Retina U-Net used a patch size of cubic 128 and a batch size of 48. These differences in training parameters reflect that ResEncL U-Net was optimized for segmentation tasks, while Retina U-Net was designed for object detection. All SSL methods used the ResEncL architecture with a patch size of 192, and a batch size of 12. 
All networks were trained for 4,000 epochs using four NVIDIA A100 GPUs and a decreasing 'poly' learning rate schedule starting at 0.01 \cite{poly_lr}. All pre-training data was preprocessed with z-score normalization and resampled to an isotropic voxel spacing of 1 mm. All other training parameters follow the default implementations in the corresponding open-source code-bases: 
All fine-tuning experiments are implemented within the nnDetection framework \cite{Baumgartner_2021,nndetv2} and follow the default training scheme, including all hyperparameters. Therefore, the computational requirements match those reported in \cite{nndetv2} and are independent of the pre-training. Supervised pre-training is conducted using the MultiTalent framework \cite{Ulrich_2023}, while self-supervised pre-training is performed using the nnSSL framework\cite{wald2024revisitingmaepretraining3d}, both inspired by nnU-Net \cite{isensee2021nnu}. 
This work establishes, for the first time, a cross-framework bridge between nnDetection and its pre-training counterparts, facilitating seamless integration across detection, segmentation and SSL paradigms. 

\subsection{Downstream Datasets}
We utilized a total of 8 datasets, comprising CT and MRI images with varying object types, to develop and evaluate all methods. The datasets were split into two pools: a development pool, which was used to determine the optimal parameters and make design decisions, and a test pool to evaluate the impact of different pre-trainings (\cref{tab:dataset_pools}). 
From all datasets without an official split we separated hold-out test sets, comprising 15-30\% of all images. The remaining images were split 80/20 into training and validation sets. Our experimental design builds upon the principles and processing steps established by Baumgartner et al. \cite{nndetv2}, ensuring consistency with their methodology. For PN9 (D07) experiments, we trained a single model on the training set, selected the post-processing parameters on the official validation set, and used the provided test set for our final evaluation. For CTA-A (D08) we split the data 80/20 into train and validation sets and utilized the internal test set (containing data from the same hospitals as the training data) for the final evaluation. To ensure a unified evaluation across the datasets, we employed the nnDetection metric calculations. For all datasets with official evaluation scripts, the official evaluation results are provided in the appendix (see section \ref{sec:official_eval}).
\begin{table}[t]
\centering
\caption{Development and Test Pool Datasets, including the numbers of images for training, validation and testing, the number of objects and the median spacing.}
\label{tab:dataset_pools}
\resizebox{\textwidth}{!}{
\begin{tabular}{llcccr}
\toprule
\textbf{Dataset} & \textbf{Target} & \textbf{Modality} & \textbf{Split} & \textbf{Objects} & \textbf{Spacing [mm]}\\
\midrule
Dev D01 MSD Pancreas \cite{MSD} & Pancreatic Tumor & CT & 156/40/85 & 283 &2.50x0.80x0.80\\
Dev D02 RibFrac \cite{ribfrac2} & Rib Fracture & CT & 336/84/80 & 4422&1.25x0.74x0.74\\
Dev D03 KiTS21 \cite{KiTS21} & Kidney Cyst, Tumor & CT & 204/51/45 & 826&0.78x0.78x0.78\\
Dev D04 LIDC \cite{LIDC}& Lung Nodule (benign vs. malign.) & CT & 690/173/155 & 1884& 1.38x0.70x0.70\\
Dev D05 DUKE Breast \cite{DukeBreast} & Primary Breast Tumor & MRI & 509/128/274 & 911&1.00x0.70x0.70\\
Test D06 LUNA16 \cite{LUNA16} & Lung Nodule & CT & 711/88/89 & 1186&1.25x0.70x0.70\\
Test D07 PN9 \cite{PN9} & Lung Nodule & CT & 6037/670/2091 & 40436  &1.00x1.00x1.00\\
Test D08 CTA-A \cite{CTA-A} & Brain Aneurysm & CT & 948/238/152  & 1590&0.40x0.46x0.46\\

\bottomrule
\end{tabular}
}
\end{table}
\subsection{Metrics and Statistical Analysis}
Detection performance was evaluated using the mean Average Precision (mAP) \cite{RetinaUNet,metrics_reloaded} at an IoU threshold of 0.1, emphasizing the diagnostic performance of the method and its ability to coarsely localize target objects. As an additional metric, the Free-response Receiver Operating Characteristic (FROC) \cite{RibFrac,LUNA16} was employed with False Positive Per Image (FPPI) thresholds at [1/8, 1/4, 1/2, 1, 2, 4, 8]. To account for variations in object counts and task difficulty, rankings were computed via bootstrapping with 1000 iterations on the image level.

\section{Experiments and Results}
\label{sec:results}
We explore large-scale pre-training for 3D object detection by evaluating four configurations:
\begin{enumerate*}[label=(\textbf{\roman*})]
\item Retina U-Net optimized for nnDetection (RetUNet),
\item Deformable DETR with the Retina U-Net encoder (DefDETR),
\item Retina U-Net with an encoder from the ResEncL architecture (ResEnc-RetUNet),
\item Deformable DETR with the ResEncL encoder (ResEnc-DefDETR).\\
\end{enumerate*}

\noindent
\textbf{Finding the Best Fine-Tuning Configuration:} \label{sec:stem} We identify the optimal fine-tuning configuration for each architecture based on MultiTalent (MT) pre-training, using an 80/20 train-validation split within the training set.
As shown in \cref{tab:finetuning}, using a fixed 1mm target spacing outperformed nnDetection’s dataset-dependent spacing on these datasets. A learning rate of 0.1 was more effective than lower values, and transferring only encoder weights performed better than full model transfer. For ResEnc, fine-tuning with the pre-training patch size (192) showed no benefit for RetUNet and caused out-of-memory issues for DefDETR on a single A100 GPU node.
Additionally, we explored strategies for handling multi-sequence datasets using D05 with four input channels. During MT pre-training, we assigned a unique stem per dataset to adjust the number of input channels, mapping them to a uniform 32-channel representation. For downstream fine-tuning, we tested three approaches:
\begin{enumerate*}[label=(\textbf{\roman*})]
    \item Random initialization,
    \item Replicating a single-channel MRI stem \cite{MSD},
    \item Using a stem from another four-sequence MRI dataset \cite{grovik2020deep}.
\end{enumerate*}
The third approach performed best. For SSL pre-training, we used the second-best random initialization instead.

\begin{table}[H]
\caption{\textbf{Finding the best fine-tuning configuration for each architecture.} Validation results on five development datasets, reporting mean Average Precision (mAP) with MultiTalent pre-training.} 
\label{tab:finetuning}
\scriptsize
\centering
\resizebox{\textwidth}{!}{
\begin{tabular}{lrrrrrccccc|ccc}\toprule
\textbf{} &\textbf{} &\textbf{} &\textbf{} &\textbf{} &\multicolumn{6}{c}{\textbf{mAP@IoU 0.1}} & \multicolumn{3}{c}{\textbf{Stem ablation D05}} \\ \cmidrule(lr){6-11} \cmidrule(lr){12-14}
\textbf{Model} &\textbf{Transfer} &\textbf{Spacing} &\textbf{Patch} &\textbf{LR} &\textbf{D01} &\textbf{D02} &\textbf{D03} &\textbf{D04} &\textbf{Mean} &\textbf{Rank} &\textbf{RND} & \textbf{1Ch}\cite{MSD}  & \textbf{4Ch}\cite{grovik2020deep}\\ \midrule
\multirow{4}{*}{\textbf{RetUNet}} &Backbone &Default & $128^3$ &1e-2 &80.32 &74.71 &80.81 &63.00 &74.71 &3.50 &- & -&-\\
&Backbone &1x1x1 &$128^3$ &1e-3 &86.96 &73.59 &79.50 &66.86 &76.73 &3.25 &- & -&-\\
&All &1x1x1 &$128^3$ &1e-2 &87.51 &\textbf{76.40} &84.04 &66.61 &78.64 &2.00  & -& -&-\\
&Backbone &1x1x1 &$128^3$ &1e-2 &\textbf{88.25} &75.74 &\textbf{85.35} &\textbf{67.32} &\textbf{79.16} &\textbf{1.25} &87.17 &84.83 &\textbf{88.16} \\\midrule
\multirow{2}{*}{\textbf{DefDETR}} &Backbone &Default &$128^3$ &3e-4 &73.32 &76.86 &\textbf{85.57} &61.96 &74.43 &1.75 &- &- &- \\
&Backbone &1x1x1 &$128^3$ &3e-4 &\textbf{90.06} &\textbf{77.82} &84.60 &\textbf{63.87} &\textbf{79.09} &\textbf{1.25} &85.96 &85.70 &\textbf{87.89}\\\midrule
\multirow{2}{*}{\shortstack[l]{\textbf{ResEnc-} \\ \textbf{RetUNet}}} &Backbone &1x1x1 &$192^3$ &1e-2 &\textbf{92.06} &73.09 &\textbf{84.23} &63.51 &78.22 &\textbf{1.50} &- &- &- \\
&Backbone &1x1x1 &$128^3$ &1e-2 &90.38 &\textbf{75.84} &83.61 &\textbf{65.96} &\textbf{78.95} &\textbf{1.50} & 84.88 &\textbf{86.12} &85.11\\ \midrule
\multirow{2}{*}{\shortstack[l]{\textbf{ResEnc-} \\ \textbf{DefDETR}}} &Backbone &1x1x1 &$192^3$ &3e-4 &OOM &OOM &OOM &OOM &- &2.00 &- &- &-\\
&Backbone &1x1x1 &$128^3$ &3e-4 &\textbf{90.53} &\textbf{77.65} &\textbf{85.49} &\textbf{66.70} &\textbf{80.09} &\textbf{1.00}  &\textbf{88.28} &85.02 &87.28\\
\bottomrule
\end{tabular}}
\end{table}

\begin{figure}[H]
\centering
\includegraphics[width=0.80\textwidth]{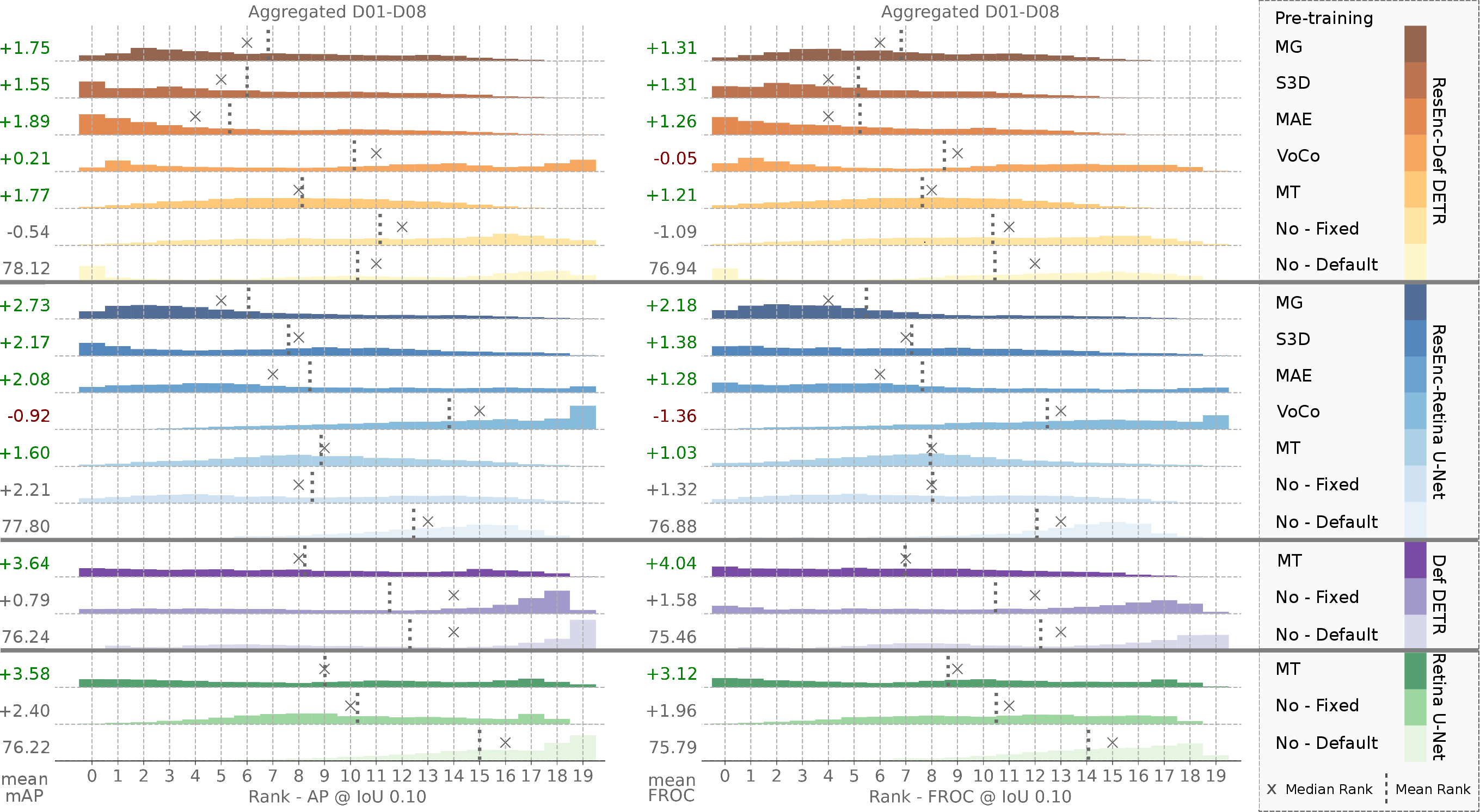}
\caption{\textbf{Reconstruction-based SSL pre-training enhances detection the most.} Aggregated ranking distributions for the test splits, that were derived from bootstrapping with 1000 iterations for each model and aggregated across all datasets D01-D08. Next to the ranking distribution, we report the difference in mAP and FROC of each method compared to the default nnDetection baseline for each architecture. %We report mean AP and FROC at an IoU threshold of 0.1.
} \label{fig:ranked_histograms}
\end{figure}

\noindent
\textbf{Impact of Pre-training}
To evaluate the impact of pre-training, we trained two baseline models from scratch for comparison: one following the architecture and configuration (e.g. median target spacing) recommended by nnDetection ("default") and another with a fixed architecture and target spacing cubic 1mm to match the pre-trained models ("fixed").
Overall, pre-trained models consistently outperform their non-pretrained counterparts across all architectures, as demonstrated in \cref{tab:comparison_pretraining} and \cref{fig:ranked_histograms}. Among the pre-training strategies, self-supervised reconstruction-based approaches (MAE, MG, S3D) yield the best results across all datasets. In contrast, contrastive pre-training (VoCo) underperforms relative to training from scratch. Supervised pre-training (MT) also leads to notable performance gains. Overall, pre-training provides a more substantial performance boost for Deformable DETR than for Retina U-Net. Furthermore, the ResEnc backbone surpasses its Retina U-Net counterpart in performance but requires more VRAM and has a higher parameter count. Notably, a fixed architecture with a target spacing of 1 mm, when trained from scratch, achieves better rankings across datasets and models than the nnDetection configuration.
\begin{table}[H]\centering
\caption{\textbf{High variance across different pre-training paradigms and architectures on the test splits of the dev. and test pool datasets.} The overall best metric is underlined, while the best for each architecture is highlighted in bold. 
}\label{tab:comparison_pretraining}
\scriptsize
\resizebox{\textwidth}{!}{
\begin{tabular}{llrrrrrrrr|rr||rrrrrrrr|rr}\toprule
\textbf{} &\multirow{2}{*}{\shortstack[l]{\textbf{Pre-}\\ \textbf{Training} \\[-5mm]}} &\multicolumn{10}{c}{\textbf{mAP@IoU 0.10}} &\multicolumn{10}{c}{\textbf{FROC@IoU 0.10}} \\
\cmidrule(lr){3-12}
\cmidrule(lr){13-22}
\textbf{Model} & &\textbf{D01} &\textbf{D02} &\textbf{D03} &\textbf{D04} &\textbf{D05} &\textbf{D06} &\textbf{D07} &\textbf{D08} &\textbf{Mean} &\textbf{Rank} &\textbf{D01} &\textbf{D02} &\textbf{D03} &\textbf{D04} &\textbf{D05} &\textbf{D06} &\textbf{D07} &\textbf{D08} &\textbf{Mean} &\textbf{Rank} \\\midrule
\multirow{3}{*}{\textbf{RetUNet}} &No - Default &73.16 &78.00 &78.38 &66.65 &78.95 &82.70 &67.12 &84.81 &76.22 &16.78 &79.50 &65.09 &73.62 &62.89 &85.40 &84.69 &65.00 &90.14 &75.79 &14.56 \\
 &No - Fixed &79.15 &\textbf{80.35} &81.40 &67.12 &75.41 &\textbf{83.33} &\textbf{68.34} &\textbf{93.89} &78.62 &9.33 &85.04 &\textbf{67.49} &76.06 &64.57 &81.49 &\textbf{84.95} &\textbf{66.38} &\textbf{96.03} &77.75 &10.11 \\
&MT \cite{Ulrich_2023} &\textbf{83.89} &79.48 &\textbf{81.83} &\textbf{68.87} &\textbf{82.44} &82.15 &67.29 &92.48 &\textbf{79.80} &\textbf{8.89} &\textbf{89.41} &66.40 &\textbf{77.48} &\textbf{65.83} &\textbf{87.90} &84.44 &65.14 &94.67 &\textbf{78.91} &\textbf{8.78} \\\midrule
\multirow{3}{*}{\textbf{DefDETR}} &No - Default &67.65 &79.93 &\textbf{83.43} &62.07 &71.06 &82.04 &\textbf{70.18} &\textbf{93.59} &76.24 &11.89 &72.61 &67.62 &77.48 &59.06 &78.47 &84.44 &67.59 &\textbf{96.37} &75.45 &11.00 \\
&No - Fixed &68.72 &80.29 &79.49 &\textbf{69.30} &74.64 &\textbf{84.31} &69.98 &89.55 &77.03 &10.89 &73.45 &68.74 &\textbf{78.25} &\underline{\textbf{68.39}} &81.18 &\textbf{86.10} &\textbf{67.73} &92.40 &77.03 &9.00 \\
&MT \cite{Ulrich_2023} &\underline{\textbf{83.96}} &\textbf{80.93} &80.61 &66.99 &\underline{\textbf{82.77}} &83.55 &69.58 &90.70 &\textbf{79.89} &\textbf{8.11} &\textbf{87.90} &\textbf{69.43} &77.99 &65.87 &\underline{\textbf{88.11}} &85.84 &67.36 &93.42 &\textbf{79.49} &\textbf{7.33} \\\midrule
\multirow{7}{*}{\shortstack[l]{\textbf{ResEnc-} \\ \textbf{RetUNet}}}&No - Default &73.84 &79.25 &79.57 &65.75 &78.64 &83.13 &68.52 &93.68 &77.80 &13.44 &78.66 &67.00 &74.39 &62.70 &84.52 &85.33 &66.20 &96.26 &76.88 &11.11 \\
&No - Fixed &\textbf{83.75} &79.35 &81.25 &68.33 &79.80 &83.69 &68.97 &94.95 &80.01 &8.00 &88.40 &67.26 &76.19 &65.50 &85.56 &86.48 &66.97 &96.60 &79.12 &7.67 \\
&MT  \cite{Ulrich_2023}&78.84 &79.70 &82.39 &67.91 &80.00 &83.45 &69.52 &93.35 &79.40 &8.11 &84.71 &66.60 &78.25 &65.87 &86.39 &85.97 &\textbf{67.69} &95.12 &78.83 &7.67 \\ &VoCo \cite{VoCo} &74.47 &\textbf{80.35} &76.00 &66.87 &78.65 &82.51 &65.86 &90.34 &76.88 &14.78 &79.66 &67.39 &72.07 &62.98 &86.44 &86.48 &63.77 &92.74 &76.44 &12.00 \\
&MAE \cite{MAE} &83.73 &78.18 &80.26 &69.07 &81.25 &83.03 &69.26 &94.27 &79.88 &8.67 &\underline{\textbf{90.25}} &63.05 &75.80 &66.29 &86.70 &\textbf{87.12} &67.27 &96.15 &79.08 &6.67 \\
&S3D \cite{tian2023designingbertconvolutionalnetworks}&79.87 &78.58 &81.10 &\underline{\textbf{70.45}} &81.24 &\textbf{83.82} &69.20 &\textbf{95.44} &79.96 &8.33 &86.39 &65.85 &76.83 &\textbf{68.30} &85.97 &86.22 &67.25 &96.60 &79.18 &7.89 \\
&MG \cite{ModelsGenesis} &82.64 &79.36 &\textbf{83.37} &68.85 &\textbf{82.06} &83.13 &\textbf{69.72} &95.05 &\underline{\textbf{80.52}} &\textbf{6.11} &87.39 &\textbf{67.65} &\underline{\textbf{79.67}} &66.39 &\textbf{87.38} &86.99 &67.60 &\textbf{96.71} &\underline{\textbf{79.97}} &\textbf{3.89} \\\midrule
\multirow{7}{*}{\shortstack[l]{\textbf{ResEnc-} \\ \textbf{DefDETR}}} &No - Default &67.87 &78.86 &\underline{\textbf{85.33}} &64.30 &80.83 &82.89 &69.32 &\underline{\textbf{95.59}} &78.12 &10.63 &73.11 &67.26 &78.89 &61.76 &85.19 &84.57 &67.01 &\underline{\textbf{97.73}} &76.94 &10.75 \\
&No - Fixed &67.20 &80.20 &80.86 &66.77 &81.60 &84.16 &69.42 &90.44 &77.58 &11.44 &73.11 &68.21 &77.61 &65.08 &86.50 &85.84 &67.50 &92.40 &77.03 &10.22 \\
&MT  \cite{Ulrich_2023}&\textbf{82.30} &81.12 &82.19 &67.09 &81.34 &83.39 &69.28 &92.47 &79.90 &7.89 &\textbf{86.89} &70.31 &78.51 &65.83 &86.18 &85.71 &67.58 &93.65 &79.33 &7.44 \\
&VoCo \cite{VoCo} &74.98 &80.22 &81.70 &63.72 &\textbf{82.72} &80.55 &70.98 &91.81 &78.34 &10.67 &79.50 &70.51 &78.51 &62.42 &\textbf{87.80} &83.04 &\underline{\textbf{68.89}} &93.88 &78.07 &8.33 \\
&MAE \cite{MAE} &74.50 &82.09 &82.21 &67.16 &81.82 &85.68 &\underline{\textbf{71.56}} &95.07 &\textbf{80.01} &\underline{\textbf{3.78}} &80.00 &69.92 &78.89 &66.15 &86.39 &87.24 &69.51 &96.94 &79.38 &3.67 \\
&S3D \cite{tian2023designingbertconvolutionalnetworks}&73.29 &\underline{\textbf{82.44}} &82.71 &\textbf{67.40} &82.14 &85.58 &70.39 &93.44 &79.67 &4.89 &80.50 &\underline{\textbf{70.61}} &\textbf{79.15}&\textbf{66.85} &87.07 &\underline{\textbf{87.50}} &68.46 &95.35 &\textbf{79.44} &\underline{\textbf{3.56}} \\
&MG \cite{ModelsGenesis} &78.62 &81.47 &82.71 &66.76 &81.41 &\underline{\textbf{85.73}} &70.44 &91.81 &79.87 &6.44 &86.05 &70.15 &78.64 &65.08 &86.44 &86.86 &68.28 &93.99 &\textbf{79.44} &6.00 \\
\bottomrule
\end{tabular}
}
\end{table}

\section{Discussion}
This work systematically studies the impact of large-scale pre-training on 3D medical object detection, showing that reconstruction-based self-supervised learning outperforms supervised pre-training. It also bridges nnDetection with pre-training frameworks, enabling a unified approach for medical image analysis. However, supervised pre-training was limited to segmentation tasks due to the scarcity of large 3D medical detection datasets, though segmentation annotations could be converted for detection. Whether organ detection pre-training truly enhances lesion detection remains questionable. Furthermore, similar to Baumgartner et al. \cite{nndetv2}, we observed high variability across tasks, which prevented us from identifying a single pre-training architecture combination that consistently outperformed all others. 
Finally, future work should investigate performance in low-data regimes and explore efficient fine-tuning strategies such as linear probing or LoRA, as these aspects were beyond the scope of this study.

\begin{credits}
\subsubsection{\ackname} This work was partially funded by the Helmholtz Foundation Model Initiative (HFMI) under the subproject The Human Radiome Project (THRP), and by Helmholtz Imaging (HI), a platform of the Helmholtz Incubator on Information and Data Science.

\subsubsection{\discintname}
The authors have no competing interests to declare that are
relevant to the content of this article. \end{credits}
%
% ---- Bibliography ----
\bibliographystyle{splncs04}
\bibliography{arXiv.bib}

\renewcommand*{\thesection}{\Alph{section}}
% \renewcommand*{\thesubsection}{\alph{subsection}.}
% \renewcommand*{\thesubsubsection}{\alph{subsubsection}.}
% \urlstyle{rm}
%
\newpage
\appendix
\section{Additional Results}

\subsection{Per Dataset Test Results}
Fig. \ref{fig:ranked_histograms_all} provides ranking histogram plots for test set results for all eight datasets. Each subplot corresponds to one dataset and shows the ranking of the evaluated models, along with the difference in mean Average Precision (mAP) compared to the nnDetection baseline. The rankings were derived from bootstrapping with 1000 iterations for each model.

By default, nnDetection automatically resamples each dataset to a target spacing, typically set to the median spacing of that dataset (\textit{default}), and selects a patch size adapted to the available GPU memory. 
In contrast, we also trained a baseline where all datasets were resampled to a fixed isotropic spacing of \mbox{$1\times1\times1\, \text{mm}^3$}, combined with a fixed patch size of \mbox{$128\times128\times128$ voxels} (\textit{fixed}). 
This configuration was chosen to match the architectural settings used during pre-training, thereby ensuring a fair comparison between baseline and pre-trained models. 
For many of the evaluated datasets, the \textit{fixed} setting improved detection performance compared to the \textit{default}, independent of any pre-training.

\subsubsection{D01 MSD Pancreas}
The MSD Pancreas task \cite{MSD,MSD2} involves the detection of pancreatic tumors. This dataset shows notable performance improvements with supervised MT pre-training, particularly for DefDETR and Retina U-Net (\cref{fig:ranked_histograms_all}). ResEncDETR also benefits from MT pre-training, while ResEnc-RetUNet shows only minor gains. In the case of ResEnc-RetUNet, a slight improvement over the \textit{fixed} architecture baseline is observed when pre-trained with Masked Autoencoder (MAE). ResEnc-DefDETR, on the other hand, consistently improves across all self-supervised learning (SSL) methods. When compared to the nnDetection \textit{default} architecture and the \textit{fixed} model, nearly all pre-trained models achieve better performance on this dataset. However, the mAP values exhibit relatively high variation across models, especially when compared to other datasets.

\subsubsection{D02 RibFrac}
\begin{figure}[H]
\centering
\includegraphics[width=0.750\textwidth]{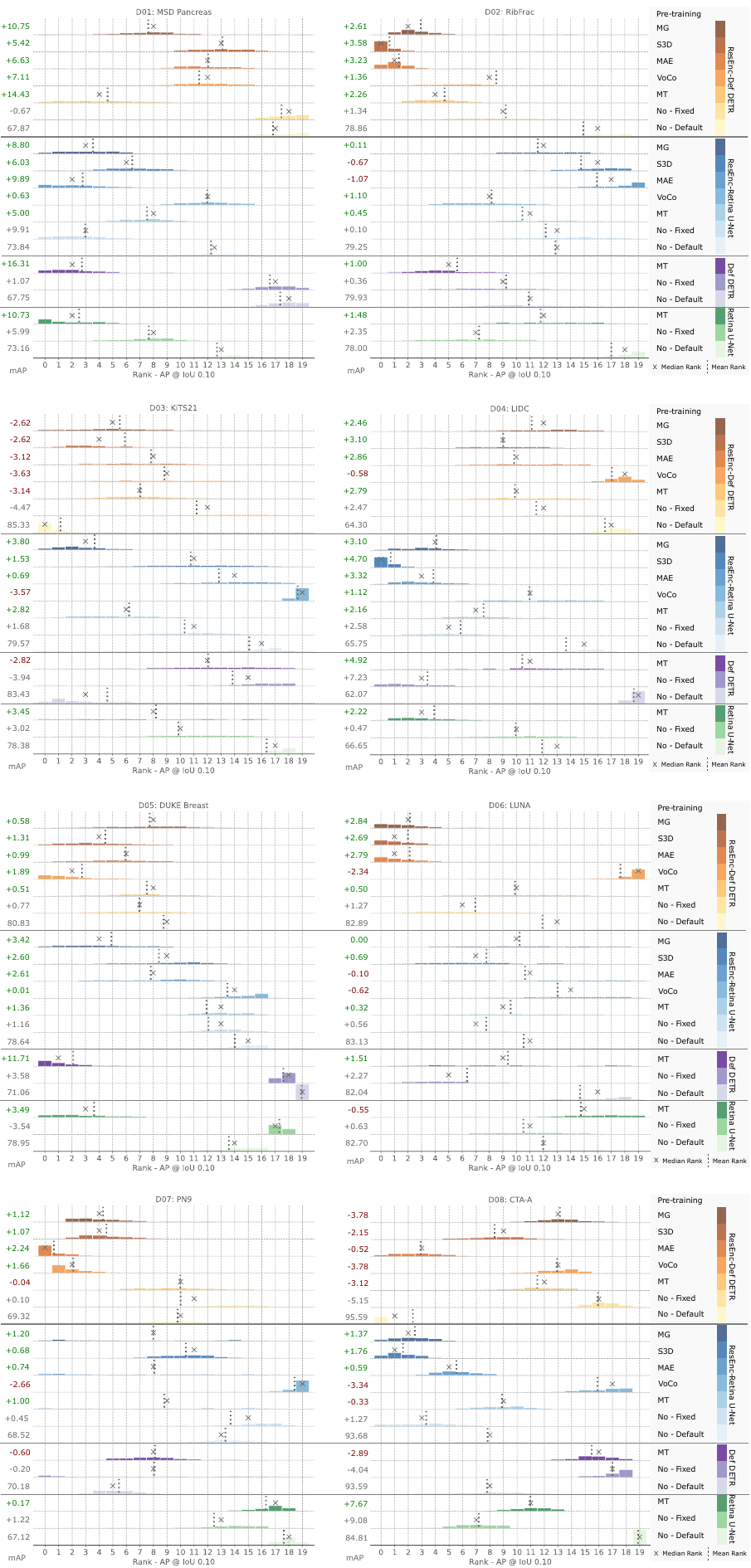}
\caption{\textbf{Per dataset test split ranking distributions.} Rankings were derived from bootstrapping with 1000 iterations for each model. Next to the ranking distribution, we report the difference in mAP of each method compared to the default nnDetection baseline for each architecture.
} \label{fig:ranked_histograms_all}
\end{figure}

The RibFrac task \cite{RibFrac} focuses on the detection of rib fractures, which differs substantially from the other datasets, as it is not centered on tumors or lesions. Moreover, the supervised pre-training dataset collection does not contain any tasks of a similar nature. Despite this, supervised pre-training still leads to performance improvements over the \textit{fixed} baseline for DETR-based architectures (\cref{fig:ranked_histograms_all}). For ResEnc-RetUNet, the effect is much smaller, and for \mbox{RetUNet}, no improvement is observed. SSL pre-training shows a performance gain for ResEnc-DefDETR, while for ResEnc-RetUNet only the contrastive method VoCo brings improvements, which is an interesting finding, since VoCo generally does not perform particularly well in our experiments. Overall, SSL pre-training outperforms supervised pre-training on this task. 

\subsubsection{D03 KiTS}
The KiTS dataset \cite{KiTS21} addresses the detection of kidney tumors and cysts. Here, the impact of pre-training differs across architectures. For RetUNet models, pre-training provides clear benefits, with supervised pre-training leading to consistent performance improvements over the \textit{fixed} baseline (\cref{fig:ranked_histograms_all}). In contrast, for DETR models, neither supervised nor SSL pre-training yields performance gains. The strong baseline performance of DETR models suggests that for KiTS the \textit{fixed} architecture may introduce a negative effect on training, thereby diminishing the potential benefits of pre-training.

\subsubsection{D04 LIDC}
The LIDC dataset \cite{LIDC} is concerned with the detection of benign and malignant lung nodules. Supervised pre-training improves performance for \mbox{RetUNet} and ResEnc-DefDETR compared to the \textit{fixed} baseline, but no such benefit is observed for other models (\cref{fig:ranked_histograms_all}). Nonetheless, both the \textit{fixed} architecture and the pre-trained models achieve better results than the nnDetection \textit{default} baseline. Among the SSL approaches, ResEnc-RetUNet with SSL pre-training performs best, and SSL pre-training also improves ResEnc-DefDETR performance. However, VoCo once again underperforms relative to other SSL methods.

\subsubsection{D05 DUKE Breast}
The DUKE Breast dataset \cite{DukeBreast} is the only MRI dataset in this study and focuses on breast tumor detection. Supervised pre-training leads to substantial performance gains for both DefDETR and RetUNet (\cref{fig:ranked_histograms_all}), despite the absence of breast-specific tasks in the MT dataset collection. For ResEnc architectures, supervised pre-training yields performance similar to the \textit{fixed} baseline. In contrast, SSL pre-training enhances performance, particularly with reconstruction-based methods, while VoCo consistently underperforms.

\subsubsection{D06 LUNA16}
The LUNA16 task \cite{LUNA16} focuses on the detection of lung nodules. Supervised MultiTalent pre-training does not appear to provide improvements over the \textit{fixed} baseline for any of the evaluated architectures (\cref{fig:ranked_histograms_all}). Nevertheless, DefDETR, ResEnc-RetUNet, and ResEnc-DefDETR with supervised pre-training still achieve higher performance than the nnDetection \textit{default} baseline. Reconstruction-based SSL pre-training yields improvements for ResEnc-DefDETR, but not for ResEnc-RetUNet. 
As in other tasks, VoCo underperforms compared to the reconstruction-based SSL pre-training strategies.

\subsubsection{D07 PN9}
The PN9 task \cite{PN9} focuses on pulmonary nodule detection and represents the largest dataset in our study. Supervised pre-training only yields a positive effect for the ResEnc-RetUNet architecture, while no benefit is observed for the other models (\cref{fig:ranked_histograms_all}). In contrast, self-supervised pre-training consistently improves performance for both ResEnc-DefDETR and ResEnc-RetUNet, with even the VoCo approach providing gains for ResEnc-DefDETR. 

\subsubsection{D08 CTA-A}
The CTA-A task \cite{CTA-A} focuses on the detection of intracranial aneurysms. 
For DETR-based architectures, enforcing the \textit{fixed} isotropic spacing and patch size leads to a degradation in performance (\cref{fig:ranked_histograms_all}). 
Although supervised pre-training, as well as all considered SSL methods, improve results compared to the \textit{fixed} baseline, they do not surpass the strong performance of the default baseline. 
It is worth noting that the supervised pre-training data did not include CTA or aneurysm datasets, but only brain imaging datasets (MRI) in general, which may explain the limited transferability. For RetUNet, supervised pre-training provides an improvement over the default baseline but does not exceed the \textit{fixed} baseline. In the case of ResEnc-RetUNet, only two reconstruction-based SSL methods achieve better performance than the \textit{fixed} baseline. MAE pre-training also improves upon the \textit{default} baseline but remains inferior to the \textit{fixed} baseline.

% \begin{figure}[H]
% \centering
% \includegraphics[width=0.750\textwidth]{fig3.pdf}
% \caption{\textbf{Per dataset test split ranking distributions.} Rankings were derived from bootstrapping with 1000 iterations for each model. Next to the ranking distribution, we report the difference in mAP of each method compared to the default nnDetection baseline for each architecture.
% } \label{fig:ranked_histograms_all}
% \end{figure}

\subsection{Comparison with nnDetectionV2 Benchmark Results}
\label{sec:official_eval}

For the test pool datasets, we followed the official evaluation protocols to compare the performance of our methods with already existing baselines and the nnDetection ensemble \cite{nndetv2}. The nnDetection ensemble typically aggregates predictions from multiple independently trained models, with different ensembles evaluated during cross-validation and the best-performing configuration selected, as reported in the nnDetectionV2 work \cite{nndetv2}. Therefore, it is not directly comparable to our single-model approaches. We nevertheless include it as a reference, since our results demonstrate that with a single pre-trained model we can approach or even exceed the performance of the ensemble, highlighting the effectiveness of pre-training for 3D medical object detection.
 According to our results on the test splits where we trained on a single fold (see \cref{fig:ranked_histograms_all}), we determined the ResEnc-DefDETR with SSL BaseMAE pre-training as our best model, which we then employed for the following experiments. 
In addition to the pre-trained model, we also trained the corresponding ResEnc-DefDETR baseline with the fixed architecture.

\subsubsection{LUNA16}
\begin{figure}[H]
\centering
\includegraphics[width=0.90\textwidth]{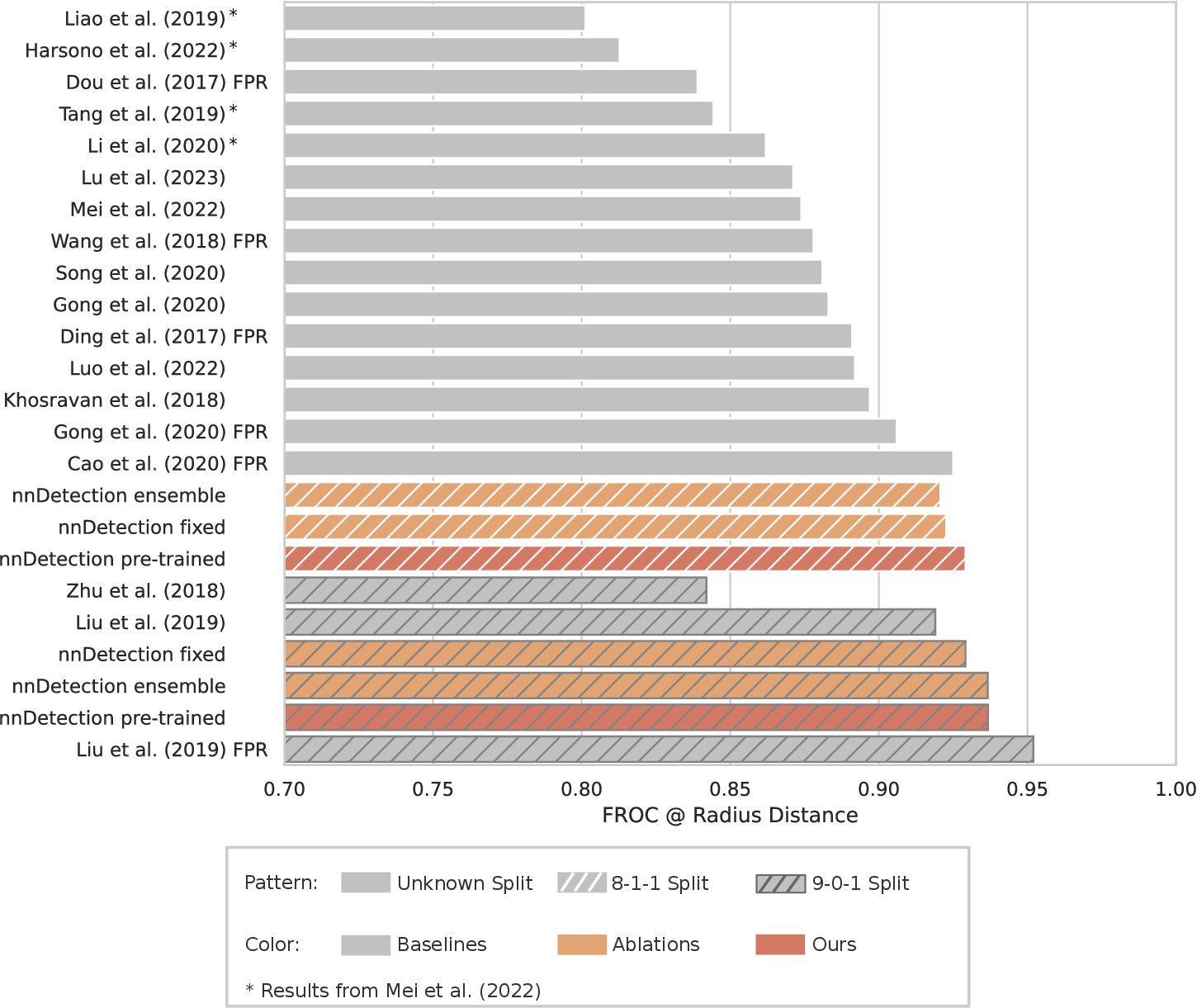}
\caption{\textbf{Comparison against LUNA16 benchmark results.} The figure shows mean FROC results for the LUNA16 dataset for 18 baseline models, the nnDetection ensemble, and our nnDetection fixed architecture without and with MAE pre-training, using two types of splits. Baseline and nnDetection ensemble results were taken directly from \cite{nndetv2}. Figure was adapted from \cite{nndetv2}.  
} \label{fig:LUNA_FROC}
\end{figure}

The official LUNA16 dataset \cite{LUNA16} consists of 888 images, divided into ten subsets, which were used for 10-fold cross-validation. Following the nnDetection work \cite{nndetv2}, we considered two splitting strategies: (i) an 8-1-1 split, where each fold was trained on eight subsets, validated on one, and tested on one; and (ii) a 9-0-1 split, where no validation set was used and training was performed on nine subsets with testing on the remaining one.
Performance was evaluated according to the official LUNA16 protocol using FROC analysis. A detection was considered a true positive if it was located within half the nodule diameter of the reference center, and the final score was defined as the mean sensitivity at seven predefined false-positive rates. Results for all 18 external baselines as well as for the nnDetection ensemble were adopted from \cite{nndetv2}.
For the 8-1-1 split, our fixed-architecture nnDetection model achieved performance comparable to the nnDetection ensemble, as shown in  \cref{fig:LUNA_FROC}. As expected, pre-training further improved performance relative to the fixed-architecture baseline. For the 9-0-1 split, the fixed-architecture model performed slightly worse than the nnDetection ensemble baseline, while the pre-trained model improved upon the fixed baseline but did not surpass the ensemble. As noted in \cite{nndetv2}, the nnDetection models are only
outperformed on the ’9-0-1 split’ by one model leveraging an additional False Positive Reduction (FPR) module.
%ensemble: split 811: RetinaNetFocalV002_D3V002Blosc_3d_split811
%split901: 'RetinaNet2SV002_D3V002Blosc_3d_split901', 'BoxDeformableDETRV002_D3V002_3d_split901'
\subsubsection{PN9}
\begin{figure}[h!]
\centering
\includegraphics[width=0.90\textwidth]{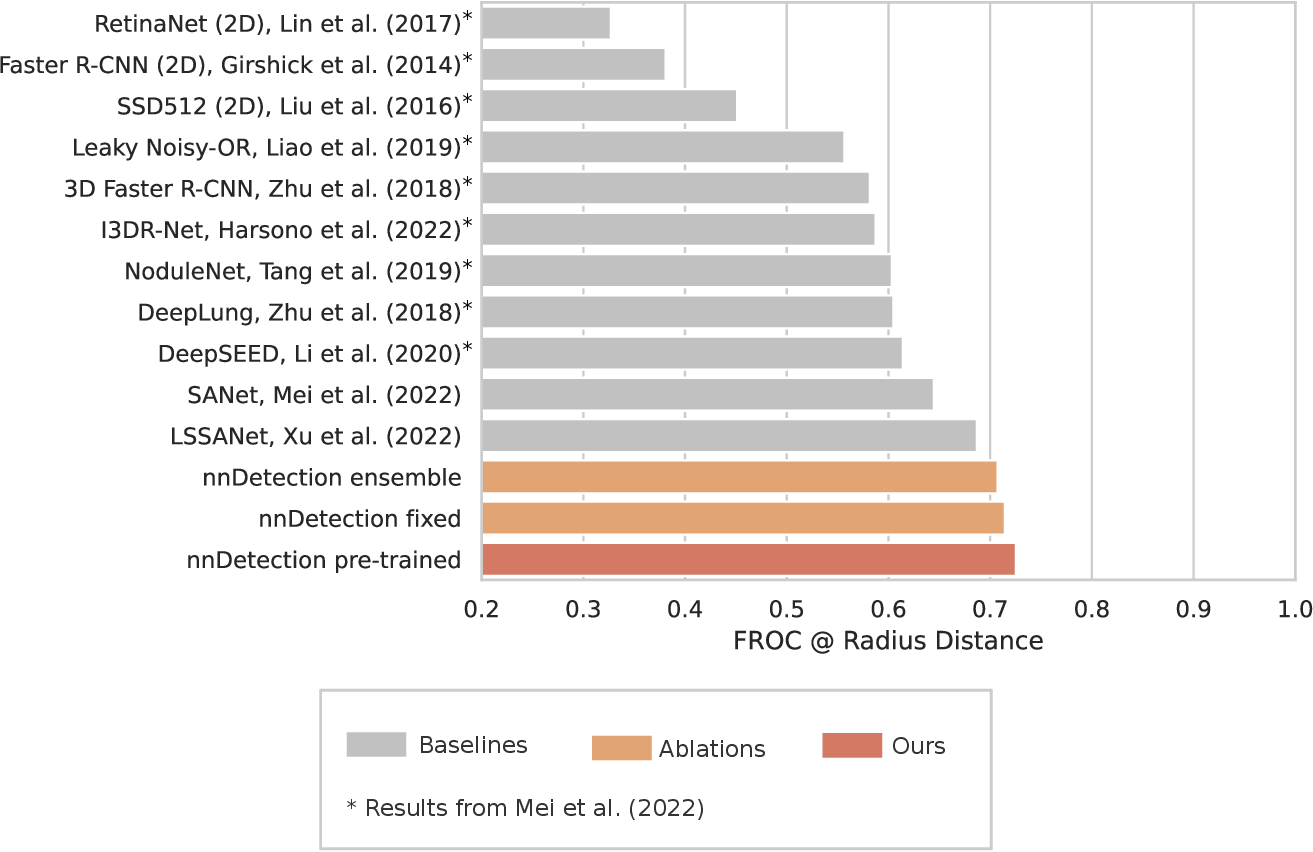}
\caption{\textbf{Comparison against PN9 benchmark results.} The figure shows mean FROC results for the PN9 dataset for 11 baseline models, the nnDetection ensemble, and our nnDetection fixed architecture without and with MAE pre-training. Baseline and nnDetection ensemble results were taken directly from \cite{nndetv2}. Figure was adapted from \cite{nndetv2}.  
} \label{fig:PN9_FROC}
\end{figure}

The PN9 dataset \cite{PN9} consists of 6,037 training images, 670 validation images, and 2,091 testing images. 
Following nnDetectionV2 \cite{nndetv2}, PN9 was trained using five-fold cross-validation, without employing the official validation set, and evaluated on the test set. 
Performance was assessed using the official PN9 FROC metric, considering
a prediction as correct if the predicted center point lies within the radius of the ground
truth object. Results for all 11 baselines and the nnDetection ensemble were adopted from \cite{nndetv2}.
As shown in \cref{fig:PN9_FROC}, all nnDetection variants achieved superior performance compared to the external baselines. Among the nnDetection configurations, the model with fixed isotropic spacing and patch size slightly outperformed the nnDetection ensemble (RetinaNet + Deformable DETR). Notably, reconstruction-based MAE pre-training further improved the FROC score, highlighting the benefit of pre-training.

\subsubsection{CTA-A}
\begin{figure}[H]
\centering
\includegraphics[width=0.80\textwidth]{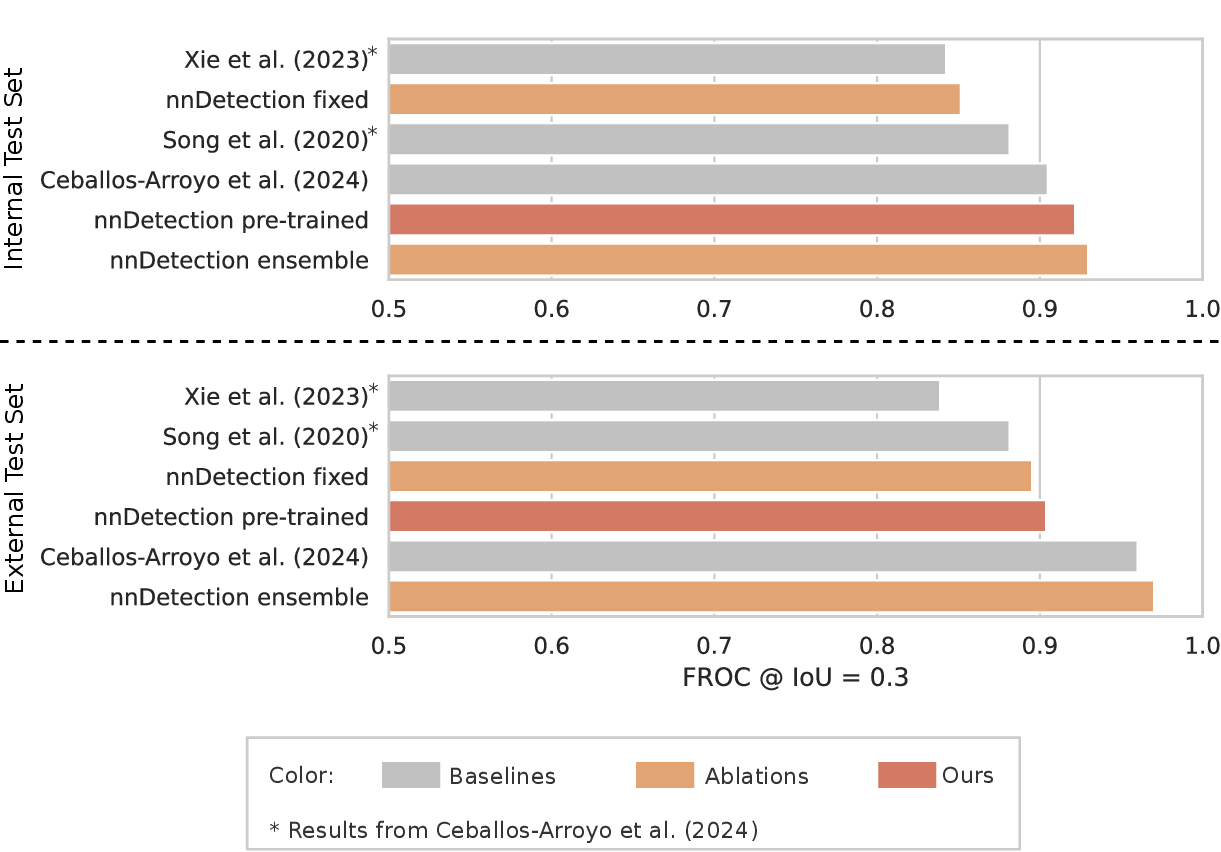}
\caption{\textbf{Comparison against CTA-A benchmark results.} The figure shows mean FROC results for the CTA-A dataset, including three baseline models, the nnDetection ensemble, and our nnDetection fixed architecture with and without MAE pre-training. Evaluation followed the official scheme on two test sets (one internal and one external). Baseline and nnDetection ensemble results were taken directly from \cite{nndetv2}. Figure was adapted from \cite{nndetv2}.
} \label{fig:CTAA_FROC}
\end{figure}

The CTA-A task \cite{CTA-A} addresses the challenging problem of intracranial aneurysm detection.  
The dataset comprises 1186 training images, for which we conducted five-fold cross-validation, and two separate test sets for final evaluation: an internal set (152 images) with a distribution similar to the training data, and an external set (138 images). The official evaluation metric is FROC at an IoU threshold of 0.3. Results for all baselines and the nnDetection ensemble were adopted from \cite{nndetv2}. On the internal test set, we observed consistent improvements with pre-training compared to training from scratch, as shown in \cref{fig:CTAA_FROC}. Our pre-trained model outperformed the external baselines, though it did not reach the performance of the nnDetection ensemble, which combines two models: a Retina U-Net and a Deformable DETR model. On the external test set, pre-training again improved performance over training from scratch, but the results remained below both the best baseline and the nnDetection ensemble.

\section{Dataset Collection for Supervised Pre-training}
\label{sec:MT_datasets}
For supervised pretraining following the MultiTalent approach \cite{Ulrich_2023}, we leveraged a large-scale and diverse collection of medical image segmentation datasets composed of 65 publicly available datasets, totaling 21,436 images across five imaging modalities: MRI (with several sequences), CT, cone-beam CT (CBCT), PET/CT, and ultrasound (US). Table \ref{tab4} provides a detailed summary of the datasets used, including dataset name, number of images, imaging modality, target anatomical structures or pathologies, and data source link. 

The dataset collection was curated to provide both anatomical diversity and clinical variability, and to not include any duplicates of images contained in the downstream detection datasets. It includes:
Organ segmentation datasets (e.g., heart: ACDC \cite{ACDC}, MSD Task 2\cite{MSD,MSD2}; liver: CHAOS \cite{CHAOSdata2019}; prostate: PROMISE12 \cite{PROMISE12}, ProstateX  \cite{prostatex}), multi-organ segmentation datasets (e.g., BTCV \cite{BTCV}, TotalSegmentator  \cite{totalseg}), lesion and tumor segmentation datasets (e.g., BraTS \cite{brats2024}, MSD-Lung \cite{MSD,MSD2}, SegTHOR \cite{lambert2019segthor}), and specialized anatomical tasks (e.g., vertebrae \cite{verse2020_1,verse2020_2,verse2020_3}, cochlear nerves \cite{crossmoda}).

Dataset sizes range from small, specialized cohorts (e.g., ACDC \cite{ACDC}: 100 cases) to large-scale, multi-center datasets (e.g., AbdomenAtlas1Mini \cite{abdomenatlas1,abdomenatlas2}: 5195 cases, AMOS22 \cite{amos}: 1055 cases).

All datasets were sourced from publicly available challenges and repositories. 
We first used this diverse collection for multi-dataset supervised pre-training of a single segmentation model. The encoder from this model was then transferred to our detection models, with the goal of leveraging the rich feature representations learned during segmentation to improve downstream 3D object detection performance.

\begin{table}[t]\centering
\caption{Description of datasets used for supervised MultiTalent pre-training. For each dataset, the table provides the dataset name, number of images, imaging modality, target anatomical structures or pathologies, and data source link.}\label{tab4}
\scriptsize
\resizebox{\textwidth}{!}{
\begin{tabular}{lrrrrr}\toprule
\textbf{Dataset} &\textbf{\# Images} &\textbf{Modalities} &\textbf{Target} &\textbf{Link} \\\midrule
Decathlon Task 2 \cite{MSD,MSD2}&20 &MRI &Heart &\url{http://medicaldecathlon.com/} \\
Decathlon Task 3 \cite{MSD,MSD2}&131 &CT &Liver &\url{http://medicaldecathlon.com/} \\
Decathlon Task 4 \cite{MSD,MSD2}&208 &MRI &Hippocampus &\url{http://medicaldecathlon.com/} \\
Decathlon Task 5 \cite{MSD,MSD2}&32 &MRI &Prostate &\url{http://medicaldecathlon.com/} \\
Decathlon Task 6 \cite{MSD,MSD2}&63 &CT &Lung &\url{http://medicaldecathlon.com/} \\
Decathlon Task 8 \cite{MSD,MSD2}&303 &CT &Hep. Vessel &\url{http://medicaldecathlon.com/} \\
Decathlon Task 9 \cite{MSD,MSD2}&41 &CT &Spleen &\url{http://medicaldecathlon.com/} \\
Decathlon Task 10 \cite{MSD,MSD2}&126 &CT &Colon &\url{http://medicaldecathlon.com/} \\
ISLES2015 \cite{isles2015}&28 &MRI &Stroke Lesion &\url{http://www.isles-challenge.org/ISLES2015/} \\
BTCV \cite{BTCV}&30 &CT &13 Abdominal Organs &\url{https://www.synapse.org/Synapse:syn3193805/wiki/89480} \\
AortaSeg24 \cite{imran2025multiclasssegmentationaorticbranches}&50 &CTA &Aorta &\url{https://aortaseg24.grand-challenge.org/} \\
AbdomenAtlas1.1Mini \cite{abdomenatlas1,abdomenatlas2}&5195 &CT &25 Abdominal Organs &\url{https://huggingface.co/datasets/AbdomenAtlas/\_AbdomenAtlas1.1Mini} \\
Promise12 \cite{PROMISE12}&50 &MRI &Prostate &\url{https://zenodo.org/records/8026660} \\
DukeLiverDatasetv2  \cite{dukeliver}&310 &MRI &Liver &\url{https://zenodo.org/records/7774566} \\
AeroPath  \cite{aeropath}&27 &CT &Lungs, Airways &\url{https://github.com/raidionics/AeroPath} \\
ACDC  \cite{ACDC}&200 &MRI &RV Cavity, Myocardium, LV Cavity &\url{https://www.creatis.insa-lyon.fr/Challenge/acdc/databases.html} \\
ISBILesion2015  \cite{ISBILesion2015}&42 &MRI &MS Lesion &\url{https://iacl.ece.jhu.edu/index.php/MSChallenge} \\
BTCV 2 \cite{BTCV2}&63 &CT &8 Abdominal Organs &\url{https://zenodo.org/records/1169361\#.YiDLFnXMJFE} \\
CHAOS  \cite{CHAOSdata2019}&60 &MRI &Liver, Kidney (L\&R), Spleen &\url{https://zenodo.org/records/3431873} \\
StructSeg Task1  \cite{structseg}&50 &CT &22 OARs (Head \& Neck) &\url{https://structseg2019.grand-challenge.org/} \\
StructSeg Task2  \cite{structseg}&50 &CT &Nasopharyngeal Cancer &\url{https://structseg2019.grand-challenge.org/} \\
StructSeg Task3  \cite{structseg}&50 &CT &6 OARs Lung &\url{https://structseg2019.grand-challenge.org/} \\
StructSeg Task4  \cite{structseg}&50 &CT &Lung Cancer &\url{https://structseg2019.grand-challenge.org/} \\
COVID-19-20 \cite{COVID19-20}&199 &CT &Lung Lesion &\url{https://covid-segmentation.grand-challenge.org/COVID-19-20/} \\
SegTHOR  \cite{lambert2019segthor}&40 &CT &Heart, Aorta, Trachea, Esophagus &\url{https://competitions.codalab.org/competitions/21145} \\
FeTA2024  \cite{FETA1,FETA2}&120 &MRI &Brain and CSF compartments &\url{https://doi.org/10.5281/zenodo.11192452} \\
ISLES2022  \cite{isles2022}&250 &MRI &Stroke Lesion &\url{https://zenodo.org/records/7153326} \\
LGGMRISeg  \cite{LGGMRISEG1,LGGMRISeg2}&110 &MRI &Glioma &\url{https://www.kaggle.com/datasets/mateuszbuda/lgg-mri-segmentation/data} \\
NIH-Pan  \cite{NIH-Pan1,NIH-Pan2}&82 &CT &Pancreas &\url{https://wiki.cancerimagingarchive.net/display/Public/Pancreas-CT} \\
M-CRIB 2.0  \cite{M-CRIB}&10 &MRI &Infant Brain Structures &\url{https://osf.io/4vthr/} \\
LVSeg  \cite{LVSeg}&1637 &MRI &Left Ventricle &\url{https://www.cardiacatlas.org/lv-segmentation-challenge/} \\
AtriaSeg  \cite{AtriaSeg}&100 &MRI &Left Atrial Cavity &\url{https://www.cardiacatlas.org/atriaseg2018-challenge/atria-seg-data/} \\
SegThyMRI\_Thyroid  \cite{thyroid}&28 &MRI &Thyroid &\url{https://www.cs.cit.tum.de/camp/publications/segthy-dataset/} \\
SegThyMRI\_all \cite{thyroid}&14 &MRI &Thyroid, Neck Vasculature &\url{https://www.cs.cit.tum.de/camp/publications/segthy-dataset/} \\
SpineMetsCTSeg \cite{spinemetseg}&55 &CT &17 Vertebrae &\url{https://www.cancerimagingarchive.net/collection/spine-mets-ct-seg/} \\
Verse2019 \cite{verse2020_2,verse2020_1,verse2020_3}&80 &CT &25 Vertebrae &\url{https://osf.io/jtfa5/} \\
Verse2020 \cite{verse2020_1,verse2020_2,verse2020_3}&61 &CT &28 Vertebrae &\url{https://osf.io/4skx2/}, \url{https://verse2020.grand-challenge.org/} \\
WMHSeg \cite{WMHSeg}&60 &MRI &White Matter Hyperintensity &\url{https://dataverse.nl/dataset.xhtml?persistentId=doi:10.34894/AECRSD} \\
BraTS2020 \cite{brats2020_1,brats2020_2,brats2020_3}&369 &MRI &Glioma &\url{https://www.kaggle.com/datasets/awsaf49/brats2020-training-data} \\
BraTS2024 Task1 \cite{brats2024,brats2024_task1} &2200 &MRI &Lower-grade Glioma &\url{https://www.synapse.org/Synapse:syn53708249/wiki/626323} \\
BraTS2024 Task2 \cite{brats2024,brats2024_task2}&60 &MRI &Glioma (African)&\url{https://www.synapse.org/Synapse:syn53708249/} \\
BraTS2024 Task3 \cite{brats2024,brats2024_task3} &500 &MRI &Meningioma &\url{https://www.synapse.org/Synapse:syn53708249/} \\
BraTS2024 Task5 \cite{brats2024,brats2024_task5} &261 &MRI &Pediatric Tumor &\url{https://www.synapse.org/Synapse:syn53708249/} \\
BraTS2024 Task6 \cite{brats2024,brats2024_task1,brats2024_task2,brats2024_task3_2023,brats2024_task4_2023,brats2024_task05_2023}&1351 &MRI &Brain Tumor &\url{https://www.synapse.org/Synapse:syn53708249/} \\
M\&Ms \cite{M&Ms1,M&Ms2}&300 &MRI &l. ventricle, r. ventricle, l. ventri. myocardium &\url{https://www.ub.edu/mnms/} \\
ToothFairy2 \cite{ToothFAiry1,ToothFairy2,ToothFairy3} &480 &CBCT &Jaw, Teeth, and Related Structures &\url{https://toothfairy2.grand-challenge.org/} \\
NCI-ISBI2013 \cite{NCI-ISBI_prostate}&59 &MRI &Prostate &\url{https://www.cancerimagingarchive.net/analysis-result/isbi-mr-prostate-2013/} \\
ProstateX  \cite{prostatex}&140 &MRI &Prostate Lesion &\url{https://www.aapm.org/GrandChallenge/PROSTATEx-2/} \\
MSLesion  \cite{MSLesion}&48 &MRI &MS Lesion &\url{https://data.mendeley.com/datasets/8bctsm8jz7/1} \\
BrainMetShare \cite{brainmetshare}&84 &MRI &Brain Metastases &\url{https://aimi.stanford.edu/datasets/brainmetshare} \\
CrossModa22 \cite{crossmoda}&168 &MRI &Vestibular Schwannoma, Cochlea &\url{https://crossmoda2022.grand-challenge.org/} \\
Atlas22 \cite{Atlas22}&655 &MRI &Stroke Lesion &\url{https://atlas.grand-challenge.org/} \\
AutoPETII \cite{autopet}&1014 &PET,CT & Whole-Body Lesions &\url{https://autopet-ii.grand-challenge.org/} \\
AMOS \cite{amos}&360 &CT,MRI &15 abdominal organs &\url{https://amos22.grand-challenge.org/} \\
TotalSegmentatorMRI \cite{totalsegmentatormri} &616 &MRI &50 classes of whole body &\url{https://zenodo.org/records/14710732} \\
TotalSegmentatorV2 \cite{totalseg}&1180 &CT &117 classes of whole body &\url{https://github.com/wasserth/TotalSegmentator} \\
HECKTOR2022 \cite{hecktor}&524 &PET,CT &Nodal Gross Tumor Volumes (Head\&Neck) &\url{https://hecktor.grand-challenge.org/} \\
MSLesionLjubljana \cite{LjubiljanaMSLesion,ljubiljanaMSLesion2}&264 &MRI &MS Lesion &\url{https://github.com/muschellij2/open\_ms\_data} \\
PENGWIN \cite{PENGWIN1,PENGWIN2} &100 &CT &Pelvic Fragments &\url{https://pengwin.grand-challenge.org/} \\
SegRap Task1 \cite{SegRap1,SegRap2}&120 &CT &45 OARs (Head\&Neck) &\url{https://segrap2023.grand-challenge.org/} \\
SegA \cite{SegA1,SegA2,SegA3}&56 &CT &Aorta &\url{https://multicenteraorta.grand-challenge.org/} \\
WORD \cite{word}&120 &CT &16 abdominal organs &\url{https://github.com/HiLab-git/WORD} \\
CTORG \cite{ctorg}&140 &CT &Lung, Brain, Bones, Liver, Kidneys and Bladder &\url{https://www.cancerimagingarchive.net/collection/ct-org/} \\
HanSeg \cite{HaNSeg}&42 &CT &30 OARs (Head\&Neck) &\url{https://han-seg2023.grand-challenge.org/} \\
TopCow \cite{topcowchallenge}&200 &CT,MRI &Vessel Components of CoW &\url{https://topcow23.grand-challenge.org/} \\\midrule
\textbf{65 Datasets} &\textbf{21,436 Imgs.} &\textbf{5 modal.} &\textbf{Various Target Structures} & \\
\bottomrule
\end{tabular}
}
\end{table}

\section{Finetuning Details}
\label{sec:finetuning}
\subsection{Preprocessing}

All downstream detection datasets underwent consistent preprocessing prior to training. Images were resampled to an isotropic voxel spacing of $1\times1\times1\,\text{mm}^3$. 

CT datasets were normalized by clipping image intensities to the 0.5th and 99.5th percentiles, followed by z-score normalization (subtracting the mean and dividing by the standard deviation). The MRI dataset D05 Duke Breast \cite{DukeBreast} was directly normalized using z-score normalization.

\subsection{Architectural Configuration and Weight Transfer}
To ensure consistency between pre-training and fine-tuning, we fixed the architectural configuration for all models. This included the number of encoder and decoder levels, the number of convolutions or residual blocks per level, feature channels, and input patch size. The respective values are listed in \cref{table_fixed_arch}.

\begin{table}[htb]
\centering
\caption{Fixed architecture configuration. The table lists the values used for spacing, patch size, strides, kernels, number of features per stage, number of convolutions per stage, and, for ResEnc models, the number of residual blocks per stage.}
\label{table_fixed_arch}

\begin{tabular}{ll}
\hline
\textbf{Parameter} & \textbf{Value} \\
\hline
Spacing & $\text{1mm} \times \text{1mm} \times \text{1mm}$ \\
Patch size & $128 \times 128 \times 128$ voxels\\
Strides & $[1,1,1], [2,2,2], [2,2,2], [2,2,2], [2,2,2], [2,2,2]$ \\
Convolutional kernels & $[3,3,3],[3,3,3],[3,3,3],[3,3,3],[3,3,3],[3,3,3]$ \\
\# Features per stage & $[32, 64, 128, 256, 320, 320]$ \\
\# Convolutions per stage & $[2,2,2,2,2,2]$ \\
\# Residual blocks per stage (ResEnc) & $[1, 3, 4, 6, 6, 6]$ \\
\hline
\end{tabular}
\end{table}

For downstream detection experiments, we transferred only the encoder weights from the pre-trained model to the downstream architecture, leaving the rest of the model weights randomly initialized. In ablation experiments, we also evaluated the effect of transferring both encoder and decoder weights for the RetUNet models, but found that transferring only the encoder yielded better performance (see \cref{sec:stem}).

For supervised MultiTalent pre-training, we transferred the single-channel stem from the TotalSegmentator CT subset to the downstream models for the single-channel CT datasets. For the multi-channel DUKE Breast MRI dataset, we systematically evaluated three options for initializing the stem: (1) random initialization, (2) replication of a single-channel stem, and (3) transfer of a pre-trained multi-channel stem (see \cref{sec:stem}). The latter option resulted in the best performance and was therefore used. 

For models pre-trained with self-supervised learning (SSL), we were unable to directly transfer pre-trained stem weights due to modality mismatches. In this case, random initialization of the stem outperformed the replication of single-channel stem weights and was thus preferred.

\subsection{Training}
Each model was trained using a train/validation/test split. The test set comprised 15–30\% of the data, while the validation set contained 20\% of the remaining training and validation pool. For the test pool datasets, we selected one of the best-performing SSL pre-trained methods, Masked Autoencoder (MAE) pre-training, and trained an additional model together with the corresponding baseline in a cross-validation fashion (10-fold for LUNA16 \cite{LUNA16} using the official 10 subsets, and 5-fold for PN9 \cite{PN9}  and CTA-A \cite{CTA-A}). We employed the official evaluation protocols for these datasets, which allowed us to directly compare our results with previously reported baselines as well as the nnDetectionv2 ensembles \cite{nndetv2}.  

For all trainings, we used the nnDetection default batch size of 4. Each epoch consisted of 2,500 training iterations and 100 validation batches. We fixed the patch size to 128x128x128 voxels and the spacing to $1\times1\times1\,\text{mm}^3$. We generally adhered to the nnDetectionV2 training schedules \cite{nndetv2} but introduced a short warm-up phase at the beginning of training to facilitate adaptation to the new tasks.

\subsubsection{RetUNet and ResEnc-RetUNet}
Both RetUNet and ResEnc-RetUNet models followed the same 50-epoch training schedule as the nnDetection Retina U-Net baseline. Training was preceded by an additional warm-up phase: during the first 5\% of iterations (7,500 iterations, corresponding to 3 epochs) only the decoder and detection heads were trained, after which the full network training schedule was applied. A linear warm-up over 4,000 iterations reached the target learning rate of $1 \times 10^{-2}$, followed by a PolyLR schedule. Optimization used stochastic gradient descent with Nesterov momentum (0.9) and a weight decay of $3 \times 10^{-5}$. The classification branch was trained with focal loss, the regression branch with L1 loss, and the segmentation branch with a combination of Dice and cross-entropy.  

RetUNet comprised approximately 19 M parameters and required about 10 GB of GPU memory, whereas ResEnc-RetUNet used the residual encoder backbone with 95.2 M parameters and required about 12 GB of memory.

\subsubsection{DefDETR and ResEnc-DefDETR}
DefDETR and ResEnc-DefDETR followed the same 100-epoch training schedule as the nnDetection DefDETR baseline (SETPREDICT model) \cite{nndetv2}, with a linear warm-up phase to a learning rate of $3 \times 10^{-4}$. During the first 5\% of iterations (12,500 iterations, 5 epochs) only the transformer and detection heads were trained. Afterwards, the full network training schedule was applied, including a 4,000-iteration linear warm-up followed by a PolyLR schedule. Optimization used Adam with AMSGrad ($\beta_1=0.9$, $\beta_2=0.999$, $\epsilon=10^{-4}$). Focal loss was used for classification, and regression was trained with a combination of L1 and GIoU losses. Auxiliary losses were calculated for each decoder block. Following \cite{nndetv2}, convergence issues occasionally observed on PN9 were addressed with an automatic restart mechanism. If model performance did not improve sufficiently within a predefined number of iterations, training was restarted with a modified warmup schedule, comprising 10,000 iterations of whole-network warmup to a lower target learning rate of $1 \times 10^{-4}$. 

DefDETR comprised about 18.2 M parameters and required about 11 GB GPU memory, while ResEnc-DefDETR comprised about 94.4 M parameters and required 15 GB GPU memory.

\section{Dataset Acknowledgements}
Data used in the preparation of this article were obtained from the Adolescent Brain Cognitive DevelopmentSM (ABCD) Study (\url{https://abcdstudy.org}), held in the NIMH Data Archive (NDA). This is a multisite, longitudinal study designed to recruit more than 10,000 children age 9-10 and follow them over 10 years into early adulthood. The ABCD Study® is supported by the National Institutes of Health and additional federal partners under award numbers U01DA041048, U01DA050989, U01DA051016, U01DA041022, U01DA051018, U01DA051037, U01DA050987, U01DA041174, U01DA041106, U01DA041117, U01DA041028, U01DA041134, U01DA050988, U01DA051039, U01DA041156, U01DA041025, U01DA041120, U01DA051038, U01DA041148, U01DA041093, U01DA041089, U24DA041123, U24DA041147. A full list of supporters is available at https://abcdstudy.org/federal-partners.html. A listing of participating sites and a complete listing of the study investigators can be found at \url{https://abcdstudy.org/consortium_members/}. ABCD consortium investigators designed and implemented the study and/or provided data but did not necessarily participate in the analysis or writing of this report. This manuscript reflects the views of the authors and may not reflect the opinions or views of the NIH or ABCD consortium investigators.

\end{document}